\newcommand{\derdef}[2]{\ensuremath{\frac{d}{d#2}#1}}

\newcommand{\tmul}[2]{\ensuremath{Tmul_{#2}(#1)}}
\newcommand{\fmul}[1]{\ensuremath{fmul(#1)}}

\newcommand{\pysmt}{\textsc{PySMT}\xspace}

\newcommand{\isatthree}{\textsc{iSAT3}\xspace}
\newcommand{\zthree}{\textsc{z3}\xspace}
\newcommand{\dreal}{\textsc{dReal}\xspace}
\newcommand{\hycomp}{\textsc{HyComp}\xspace}
\newcommand{\hyst}{\textsc{HYST}\xspace}
\newcommand{\nuxmv}{\textsc{nuXmv}\xspace}

\newcommand{\dreach}{\textsc{dReach}\xspace}

\newcommand{\icthreeia}{\textsc{IC3}\xspace}
\newcommand{\bmc}{BMC\xspace}
\newcommand{\kinduction}{k-induction\xspace}

\newenvironment{small2}{\fontsize{8}{10}\selectfont}{\normalsize}
\newsavebox{\fmbox}
\newenvironment{fmpage}[1]
{\begin{lrbox}{\fmbox}\begin{minipage}{#1}}
{\end{minipage}\end{lrbox}\fbox{\usebox{\fmbox}}}
\newcommand{\defas}{\ensuremath{\stackrel{\text{\tiny def}}{=}}\xspace}
\newcommand{\abst}[1]{\ensuremath{\widehat{#1}}}

\newcommand{\UFLRA}{\text{LRA+EUF}\xspace}
\newcommand{\NRA}{\text{NRA}\xspace}
\newcommand{\LRA}{\text{LRA}\xspace}
\newcommand{\functionfont}[1]{\textsf{\scalebox{0.9}{#1}}}
\newcommand{\SMTinitialabstraction}{\ensuremath{\functionfont{initial-abstraction}}\xspace}
\newcommand{\SMTgetNRAmodel}{\ensuremath{\functionfont{get-\NRA-model}}\xspace}
\newcommand{\SMTrefine}{\ensuremath{\functionfont{refine}}\xspace}
\newcommand{\SMTite}{\ensuremath{\functionfont{ite}}\xspace}
\newcommand{\SMTconcretize}{\ensuremath{\functionfont{concretize}}\xspace}
\newcommand{\SMTNRAcheck}{\ensuremath{\functionfont{SMT-\NRA-check}}\xspace}
\newcommand{\SMTNRAcheckCEGAR}{\ensuremath{\functionfont{SMT-\NRA-check-abstract}}\xspace}
\newcommand{\SMTNRAcheckCEGARext}{\ensuremath{\functionfont{SMT-\NRA-check-abstract-ext}}\xspace}
\newcommand{\SMTUFLRAcheck}{\ensuremath{\functionfont{SMT-\UFLRA-check}}\xspace}
\newcommand{\mktuple}[1]{\ensuremath{\langle #1\rangle}\xspace}

\newcommand{\VMTinitialabstraction}{\ensuremath{\functionfont{initial-abstraction}}\xspace}
\newcommand{\VMTNRAcheckname}{IC3-\NRA-prove}
\newcommand{\VMTNRAcheck}{\ensuremath{\functionfont{\VMTNRAcheckname}}\xspace}
\newcommand{\VMTUFLRAcheck}{\ensuremath{\functionfont{IC3-\UFLRA-prove}}\xspace}
\newcommand{\VMTgetcexformula}{\ensuremath{\functionfont{get-cex-formula}}\xspace}
\newcommand{\VMTrefine}{\ensuremath{\functionfont{refine-transition-system}}\xspace}
\newcommand{\VMTreduceaxioms}{\ensuremath{\functionfont{reduce-axioms}}\xspace}
\newcommand{\TS}{{\cal S}\xspace}

\newcommand{\ourapproach}{\VMTNRAcheckname\xspace}
\newcommand{\isatdefault}{iSAT3[1e-1]\xspace}
\newcommand{\isatprecision}{iSAT3[1e-9]\xspace}
\newcommand{\staticaxioms}{\nuxmv-LRA-static\xspace}
\newcommand{\nrabmc}{\NRA-BMC}
\newcommand{\bmczthree}{\nrabmc-Z3\xspace}
\newcommand{\bmcdreal}{\nrabmc-DREAL\xspace}
\newcommand{\nrakind}{\NRA-K-induction}
\newcommand{\kindzthree}{\nrakind-Z3\xspace}
\newcommand{\kinddreal}{\nrakind-DREAL\xspace}

\newcommand{\varsof}[1]{\ensuremath{\text{vars}(#1)}\xspace}
\newcommand{\atomsof}[1]{\ensuremath{\text{atoms}(#1)}\xspace}
\newcommand{\mksubst}[3]{\mbox{\ensuremath{#1{\{#2 \mapsto #3\}}}}\xspace}
\newcommand{\attime}[2]{\ensuremath{#1}^{\mktuple{#2}}}

\newcommand{\vi}{\ensuremath{\varphi}}
\newcommand{\sref}[1]{\S{}\ref{#1}}
\newcommand{\SMTgetassignment}{\ensuremath{\functionfont{get-assignment}}\xspace}
\newcommand{\SMTmultiplicationlines}{\ensuremath{\functionfont{linearization-axioms}}\xspace}

\documentclass[a4paper]{llncs}
\usepackage{makeidx}  %
\usepackage[pdftex]{graphicx}
\usepackage{amsmath}
\usepackage{amsfonts}
\usepackage{commath}
\usepackage{listings}
\usepackage{cite}
\usepackage{times}
\usepackage{placeins} %
\usepackage{xspace}
\usepackage{tabularx}
\usepackage{multirow}
\usepackage[caption=false,font=footnotesize]{subfig}

\usepackage{todonotes}

\begin{document}

\title{%
{Invariant Checking of NRA Transition Systems\\
via Incremental Reduction to LRA with EUF\thanks{This work was performed as part of the H2020-FETOPEN-2016-2017-CSA project SC$^2$ (712689).
}}}
\titlerunning{CEGAR VMT}  %
\author{Alessandro Cimatti\inst{1}
\and Alberto Griggio\inst{1} \and Ahmed Irfan\inst{1,2} \and \\ Marco Roveri\inst{1} \and Roberto Sebastiani\inst{2}}
\institute{Fondazione Bruno Kessler, Italy,\\
\email{\texttt{[lastname]}@fbk.eu},\\
\and
University of Trento, Italy,\\
\email{\texttt{[firstname].[lastname]}@unitn.it}
}

\maketitle              %

\sloppypar

\begin{abstract}
Model checking invariant properties of designs, represented as
transition systems, with non-linear real arithmetic (NRA), is an
important though very hard problem. On the one hand NRA is a
hard-to-solve theory; on the other hand most of the powerful model
checking techniques lack support for NRA. In this paper, we present a
counterexample-guided abstraction refinement (CEGAR)
approach that %
leverages linearization techniques from differential calculus to
enable the use of mature and efficient model checking algorithms %
for
transition systems on linear real arithmetic (LRA) with uninterpreted functions (EUF).
The results of an empirical evaluation  confirm the validity and potential of this approach.
\end{abstract}

\section{Introduction}
\label{sec:introduction}

Invariant checking for infinite-state transition systems is a
fundamental research area. Based on the recent improvements of SMT
technologies, effective approaches have been developed for the case of
transition systems with dynamics over Linear Real
Arithmetic~\cite{ic3-ia,gpdr,spacer,ctigar}.
However, many real-world industrial designs (e.g.  aerospace,
automotive) require modeling as transition systems over non-linear
arithmetic (NRA).
Although both problems are undecidable, proving properties of the NRA
transition systems turns out to be much harder than the linear case,
and has in fact received much less attention.
Approaches based on BMC and k-induction~\cite{k-induction,temporal-induction} are possible, so that
non-linearity is handled at the SMT-level, by means of an SMT(NRA)
solver (e.g. Z3~\cite{z3}, nlSAT~\cite{jovanovic2012solving},
Yices~\cite{yices}, SMT-RAT~\cite{abraham2010lazy}).
Their power is however limited. Consider the following simple
transition system: initially, $x \geq 2 \wedge y \geq 2 \wedge z =
x*y$; the transition relation is defined by $x'= x + 1 \wedge y' = y +
1 \wedge z' = x'*y'$.  The property ``{\em it is always the case that
  $z \geq x + y$}'' is not k-inductive, not even for a very large
value of k. Thus, the typical proving techniques that are based
on k-induction using an SMT(NRA) solver will not be able to prove it.
In principle, it is also possible to lift other approaches
(e.g. interpolation, IC3~\cite{interpolation-mc,ic3-ia}) to handle
non-linearities at the level of the solver. However, this requires the
SMT(NRA) solver to carry out interpolation or quantifier elimination,
and to proceed incrementally. These extra functions are usually not
available, or they have a very high computational cost.

In this paper, we propose a completely different approach to tackle
invariant checking for NRA transition systems. Basically, we work with
an abstract version of the transition system, expressed over LRA with
EUF, for which we have effective verification tools~\cite{ic3-ia}.
In the abstract space, nonlinear multiplication is modeled as an
uninterpreted function. When spurious counter-examples are found, the
abstraction is tightened by the incremental introduction of linear
constraints, including tangent planes resulting from differential
calculus, and monotonicity constraints.

We implemented the approach on top of the \nuxmv model
checker~\cite{nuxmv}, leveraging the IC3 engine with Implicit Abstraction~\cite{ic3-ia} for invariant
checking of transition systems over LRA with EUF.
We compared it, on a wide set of benchmarks, against multiple
approaches working at NRA level, including BMC and k-induction using
SMT(NRA), the recent interpolation-based \isatthree engine~\cite{isat-hvc}, and
the static abstraction approach proposed in~\cite{ChampionGKT16}. The
results demonstrate substantial superiority of our approach, that is
able to solve the highest number of benchmarks.

The effectiveness of our approach is possibly explained with the
following insights. On the one hand, in contrast to LRA, NRA is a
hard-to-solve theory: in practice, most available complete solvers rely on CAD
techniques~\cite{Collins1974_CAD}, which require double exponential
time in worst case. Thus, we try to avoid NRA reasoning, trading it
for LRA and EUF reasoning.
On the other hand, proving properties of practical NRA transition
systems may not require the full power of non-linear solving. In fact,
some systems are ``mostly-linear'' (i.e.  non-linear constraints are
associated to a very small part of the system), an example being the Transport Class Model (TCM) for aircraft simulation from the Simulink model library~\cite{hueschen2011development}.
Furthermore, even NRA transition systems with significant non-linear
dynamics may admit a piecewise-linear invariant of the transition
system that is strong enough to prove the property.

\emph{Structure.} In Sec.~\ref{sec:related-works} we discuss the
related work, and in Sec.~\ref{sec:background} introduce some
background.  In Sec.~\ref{sec:cegar-nra-smt} we discuss the approach in
the setting of SMT(NRA). In Sec.~\ref{sec:cegar-nra} we present the
verification algorithm for NRA transition systems. In
Sec.~\ref{sec:experiments} we describe the results of the experimental
evaluation.  In Sec.~\ref{sec:conclusion} we conclude and outline the
directions for future research.

\section{Related Work}
\label{sec:related-works}

There are not many tools that deal with NRA transition systems.
The most relevant is the recently proposed
\isatthree~\cite{scheibler2013recent}, that uses an
interpolation-based~\cite{kupferschmid2011craig, isat-hvc} approach to
prove invariants. In addition to NRA, it also supports trascendental
functions and some form of differential equations.  \isatthree is
built on an SMT solver based on numeric techniques (interval
arithmetic), and is able to provide results that are accurate up to
the specified precision. In fact, in addition to ``safe'' and
``unsafe'' answers, \isatthree may return ``maybe unsafe'' when it
finds an envelope of given precision that may (but is not guaranteed
to) contain a counterexample.
Another relevant tool is \dreach~\cite{kong2015dreach}, a bounded
model checker implemented on top of the \dreal~\cite{gao2013dreal} SMT
solver, that adopts numerical techniques similar to
\isatthree. \dreach has an expressiveness similar to \isatthree, but
being a bounded model checker it is unable to prove properties.

The work in \cite{ChampionGKT16} follows a reduction-based approach to
check invariants of \NRA transition systems. It over-approximates the
non-linear terms with a coarse abstraction, encoding into \LRA some weak
properties of multiplication like identity and sign.
Another reduction-based approach is presented
in~\cite{marechal2016polyhedral} in the context of program
analysis. The idea is to find a (tighter) convex
approximation of polynomials in form of polyhedra, thus obtaining a
conservative linear transition system.
The key differences of our approach with respect
to~\cite{ChampionGKT16,marechal2016polyhedral} are that we iteratively
refine the abstraction, and we adopt a reduction to \UFLRA.
Furthermore, to the best of our knowledge, there is no available
implementation of the approach~\cite{marechal2016polyhedral} in a
program analysis tool -- it has been only shown to work on SMT
problems.

The idea of approximating a univariate function (in particular the
natural logarithm $ln$) with tangent lines is used
in~\cite{tiwari2015time}. Here we abstract a bivariate function
(multiplication), and use tangent planes for the refinement. We
also exploit other properties (e.g. monotonicity) to derive additional axioms.
The idea of using tangent planes (spaces) has been explored
in~\cite{nuzzo2010calcs}, limited to the case of SMT
solving. Another key differences is that the tangent planes area used
to under-approximate predicates, while we use them to refine the
over-approximation of the multiplication function.

\section{Background}
\label{sec:background}

\subsubsection{Properties of the Multiplication Function.}
\label{subsec:calculus}

\begin{figure*}[!t]
\centering
\subfloat[$x*y$]{\includegraphics[scale=.15]{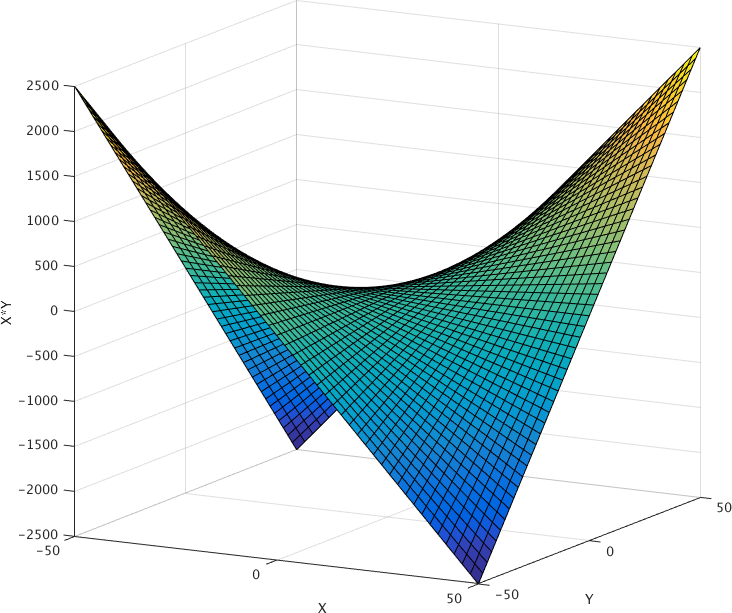}%
\label{fig:mul-2var}
}~
\subfloat[$x*y$ (top view)]{\includegraphics[scale=.12]{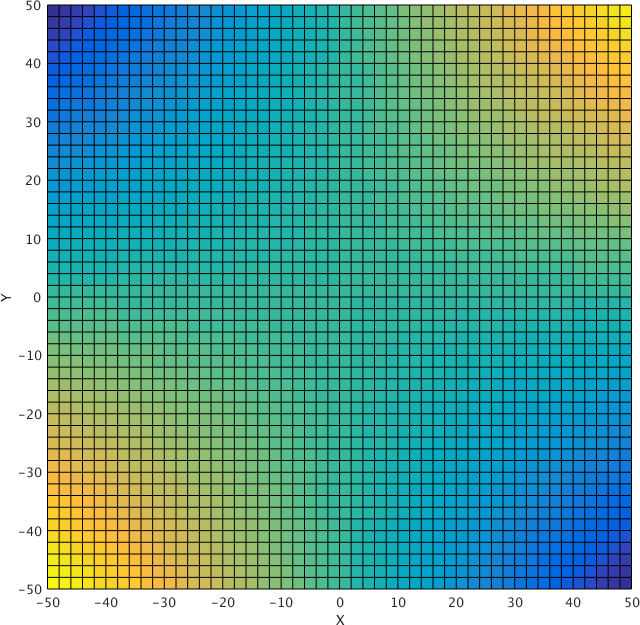}%
\label{fig:mul-2var-top}
}~
\subfloat[$x*y$ and tangent plane]{\includegraphics[scale=.15]{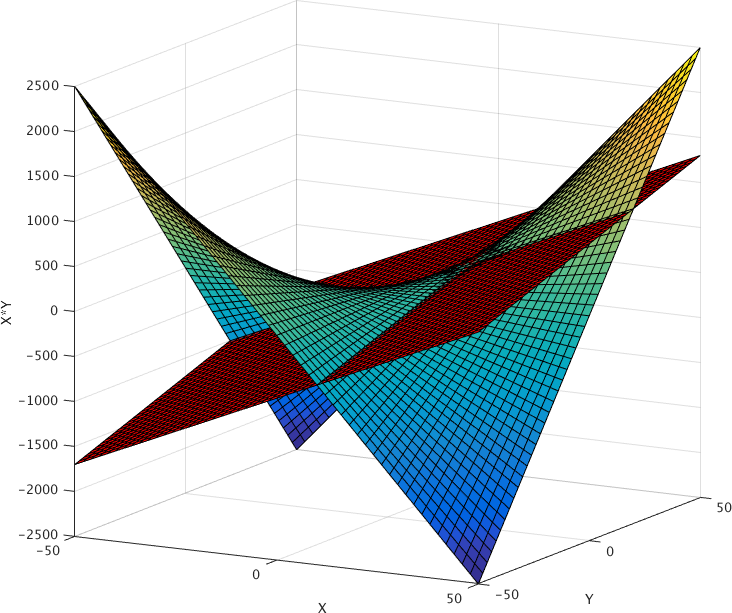}%
\label{fig:mul-2var-tangent}
}~
\subfloat[$x*y$ and tangent plane (top view)]{\includegraphics[scale=.12]{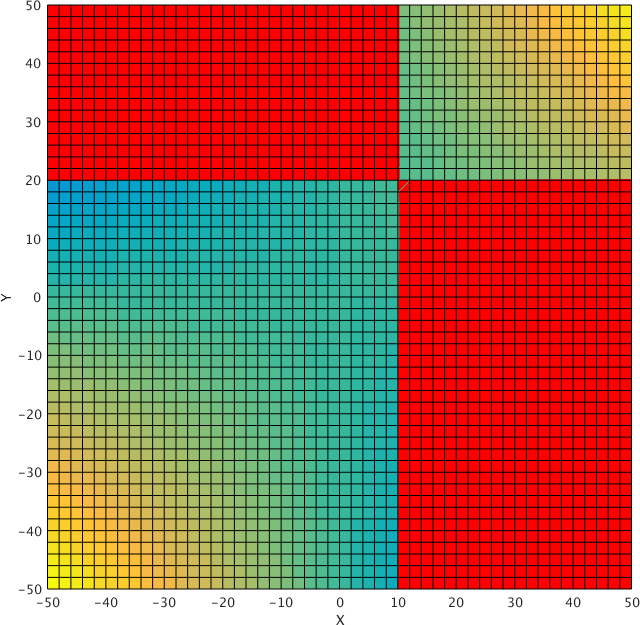}%
\label{fig:mul-2var-tangent-top}
}
\caption{Multiplication function and tangent plane.
  \label{fig:plots-mult-tangent}}
\end{figure*}

Geometrically, the surface generated by the multiplication function 
$f(x,y) \defas x * y$ is shown in Fig.~\ref{fig:mul-2var}~and~\ref{fig:mul-2var-top}.
This kind of surface is known in geometry as hyperbolic paraboloid.
A hyperbolic paraboloid is a doubly-ruled surface, i.e. for every
point on the surface, there are two distinct lines projected from the
surface such that they pass through the point. In case of the
multiplication surface, the projected lines basically lie on the
surface.

\paragraph{Tangent Plane.}
The tangent plane to a surface at a point of interest $(a,b)$ is a
plane that ``just touches'' the surface at the point. The tangent
planes can be used to linearly approximate the surface at the point of
interest. An important property of the tangent plane to a hyperbolic
paraboliod is that the two projected lines from the surface are also
in the tangent plane, and they define how the plane cuts the surface
(see
Fig.~\ref{fig:mul-2var-tangent}~and~\ref{fig:mul-2var-tangent-top}).
The tangent plane $\tmul{x,y}{a,b}$ to the multiplication function
$f(x,y)$ at point $(a,b)$ is calculated as follows:
\begin{align*}
\begin{split}
  \tmul{x,y}{a,b} \defas f(a,b) & +  \derdef{f(x,y)}{x}_{\mid_{(a,b)}}*(x - a) +  \derdef{f(x,y)}{y}_{\mid_{(a,b)}}*(y - b)
\end{split}
\end{align*}
where $\derdef{f(x,y)}{x}_{\mid_{(a,b)}}$ and $\derdef{f(x,y)}{y}_{\mid_{(a,b)}}$ are the first-order partial
derivatives of $f(x,y)$ w.r.t. $x$ and $y$ respectively, evaluated at $(a,b)$.
$\tmul{x,y}{a,b}$ simplifies to:
\begin{equation}
\tmul{x,y}{a,b} \defas b*x + a*y - a*b
\label{eq:mult-tangent-plane}
\end{equation}

\subsubsection{Logic and Satisfiability.}

We assume the standard first-order quantifier-free logical setting and
standard notions of theory, model, satisfiability, and logical consequence. 
If $\varphi$ is a formula, 
we denote with $\varsof{\varphi}$ the set of its variables,
and with $\atomsof{\varphi}$ the set of its atoms.
We write $\varphi(X)$ to denote that $\varsof{\varphi} \subseteq X$.
If $x$ and $y$ are two variables, we denote with $\mksubst{\varphi}{x}{y}$ 
the formula obtained by replacing all the occurrences of $x$ in $\varphi$ with $y$. 
We extend this notation to ordered sequences of variables in the natural way.
If $\mu$ is a model and $x$ is a variable, we write $\mu[x]$ to denote the value of $x$ in $\mu$, 
and we extend this notation to terms in the usual way.
If $X$ is a set of variables, 
we denote with $X'$ the set obtained by replacing 
each element $x \in X$ with $x'$, 
and with $\attime{X}{i}$ the set obtained by replacing $x$ with $\attime{x}{i}$.
If $\Gamma$ is a set of formulas, 
we write $\bigwedge\Gamma$ to denote the formula obtained by taking the conjunction of all its elements.
If $\bigwedge\Gamma$ is unsatisfiable (modulo some theory $T$), 
an unsatisfiable core is a set $C \subseteq \Gamma$ such that $\bigwedge C$ is still unsatisfiable.

\subsubsection{Symbolic Transition Systems.}
\label{subsec:transition-system}

A symbolic transition system $\TS \defas \mktuple{X, I, T}$ 
is a tuple where $X$ is a finite set of (state) variables,
$I(X)$ is a formula denoting the initial states of the system,
and $T(X,X')$ is a formula expressing its transition relation.
A state $s_i$ of $\TS$ is an assignment to the variables $X$.
A path (execution trace) $\pi = s_0,s_1,s_2,\ldots,s_{k-1}$ of length $k$ 
(possibly infinite)
for $\TS$ is a sequence of states 
such that $s_0 \models I$ and 
$s_i \land \mksubst{s_{i+1}}{X}{X'} \models T$ for all $0 \leq i < k-2$.
We call an unrolling of $\TS$ of length $k$ the formula 
$\mksubst{I}{X}{\attime{X}{0}} \land \bigwedge_{i=0}^{k-1}\mksubst{\mksubst{T}{X}{\attime{X}{i}}}{X'}{\attime{X}{i+1}}.$

Let $P(X)$ be a formula whose assignments represent a property (good
states) over the state variables $X$. 
The invariant verification problem,
denoted with $S\models P$, is the problem of checking if for all the
finite paths
$s_0,s_1,\ldots,s_k$ of $\TS$, for all $i$, $0\leq i\leq k$,
$s_i\models P$.
Its dual formulation in terms of reachability of $\neg P$ is the
problem of finding a path $s_0,s_1,\ldots,s_k$ of $\TS$ such that
$s_k\models \neg P$. $P$ represents the ``good'' states, while
$\neg P$ represents the ``bad'' states.

\section{Solving SMT(\NRA) via SMT(\UFLRA)}
\label{sec:cegar-nra-smt}

\subsubsection{Top-level Algorithm.}

The main idea of this paper is that of solving an SMT formula
containing non-linear polynomial constraints (i.e., expressed in the
\NRA theory) by overapproximating it with a formula over the combined
theory of linear arithmetic and uninterpreted functions (\UFLRA).
Our main SMT solving procedure follows a classic abstraction refinement loop,
in which at each iteration the current overapproximation of the input SMT formula
is refined by adding new constraints that rule out one (or possibly more) spurious solutions,
until one of the following occurs:
(i) the SMT formula becomes unsatisfiable in the \UFLRA theory; or
(ii) the \UFLRA model for the current overapproximation can be lifted to an \NRA model for the original SMT formula; or
(iii) the resource budget (e.g. time, memory, number of iterations) is exhausted.

The pseudocode for the top-level algorithm is shown in Fig.~\ref{fig:smt-nra-pseudocode}.
We provide more details about its main components in the rest of this section.

\newcounter{pseudocodecounter}
\newcommand{\pcl}{%
  \refstepcounter{pseudocodecounter}
  {\scriptsize \arabic{pseudocodecounter}.}\hspace{1em}%
}
\newcommand{\pcll}[1]{\pcl\label{#1}} %
\newcommand{\pcc}[1]{%
  {\it ~\# #1}%
}
\newcommand{\pcs}{%
  {\phantom {\scriptsize \arabic{pseudocodecounter}.}\hspace{1em}}
}

\begin{figure}[t]
  \centering
    \begin{fmpage}{1.0\linewidth}
  \begin{small2}
    \setcounter{pseudocodecounter}{0}
      \begin{tabbing}
        {\bf bool} \SMTNRAcheckCEGAR($\varphi$):\\
        \pcs xxx \= xxx \= xxx \= xxx \= xxx \= xxx \= \kill
        \pcll{code:initial-abstraction} $\abst{\varphi} = \SMTinitialabstraction(\varphi)$\\
        \pcl res, data = $\SMTNRAcheckCEGARext(\abst{\varphi})$ \\
        \pcl {\bf return} res
      \end{tabbing}
      \begin{tabbing}
        $\langle${\bf bool}, axiom set or model$\rangle$ \SMTNRAcheckCEGARext($\abst{\varphi}$):\\
        \pcs xxx \= xxx \= xxx \= xxx \= xxx \= xxx \= \kill
        \pcl $\Gamma = \emptyset$\\
        \pcl {\bf while} true:\\
        \pcl \> {\bf if} budget-exhausted(): {\bf abort}\\
        \pcll{code:UF-LRA-check} \> res, $\abst{\mu} = \SMTUFLRAcheck(\abst{\varphi} \land \bigwedge \Gamma)$ \\
        \pcl \> {\bf if not} res: {\bf return} false, $\Gamma$\\
        \pcll{code:get-NRA-model} \> ok, $\mu = \SMTgetNRAmodel(\abst{\varphi}, \abst{\mu})$\\
        \pcl \> {\bf if} ok: {\bf return} true, $\mu$\\
        \pcll{code:refine} \> {\bf else}: $\Gamma = \Gamma \cup \SMTrefine(\abst{\mu})$
      \end{tabbing}
  \end{small2}
    \end{fmpage}
  \caption{Solving SMT(\NRA) via abstraction to SMT(\UFLRA).
    \label{fig:smt-nra-pseudocode}}
\end{figure}

\subsubsection{Initial Abstraction.}
\label{sec:smt-initial-abstraction}

The function \SMTinitialabstraction takes as input an SMT(\NRA) formula $\varphi$
and returns an overapproximation $\abst{\varphi}$ of it in the \UFLRA theory.

First, each multiplication expression $x * y$
between two variables%
\footnote{To simplify the presentation, we assume (here and in the rest of the paper)
that all multiplications in $\varphi$
are either between two variables or between one constant and one variable.
}
occurring in $\varphi$ is replaced by $\fmul{x,y}$,
where $\fmul{}$ is a binary uninterpreted function returning a real.
We remark that this happens only for non-linear multiplications:
expressions like $c * x$ or $x * c$ in which $c$ is a constant are not rewritten.

Then, some simple axioms about multiplication are added to $\abst{\varphi}$
via static learning. For each $\fmul{x,y} \in \abst{\varphi}$, we add the following axioms:
\begin{align*}
\textbf{Commutativity: } & \fmul{x,y} = \fmul{y,x}\\
\textbf{Sign: }          & \fmul{x,y} = \fmul{-x,-y} \wedge \fmul{x,y} = -\fmul{-x,y} ~\land\\
                         & \fmul{x,y} = -\fmul{x,-y}\\
\textbf{Zero: }          & ((x = 0 \lor y = 0) \leftrightarrow \fmul{x,y}=0) ~\land\\
                         & (((x > 0 \wedge y > 0) \vee (x < 0 \wedge y < 0)) \rightarrow \fmul{x,y} > 0) ~\land\\
                         & (((x < 0 \wedge y > 0) \vee (x < 0 \wedge y > 0)) \rightarrow \fmul{x,y} < 0)
\end{align*}

\subsubsection{Abstraction Refinement.}
\label{sec:smt-abstractionrefinement}

If the SMT check on the \UFLRA abstraction returns false
(line~\ref{code:UF-LRA-check} of Fig.~\ref{fig:smt-nra-pseudocode}),
we can conclude that the input formula is unsatisfiable. In this case,
$\Gamma$ contains all the lemmas (discussed later in this section)
that were added in the earlier refinements (line~\ref{code:refine} of
Fig.~\ref{fig:smt-nra-pseudocode}).

Otherwise, we have to check whether the model $\abst{\mu}$
found for $\abst{\varphi}$ is also a model for the original
\NRA formula $\varphi$.
Let $Fmuls$ be the set of all $\fmul{x,y}$ terms occurring in $\abst{\varphi}$.
In its simplest version,
the function \SMTgetNRAmodel checks whether, for all $\fmul{x,y}$ in $Fmuls$,
$\abst{\mu}[\fmul{x,y}] = \abst{\mu}[x] * \abst{\mu}[y]$.
If this is the case, then $\abst{\mu}$ is also a model for the original formula,
and \SMTgetNRAmodel returns true.
(We present more sophisticated versions of \SMTgetNRAmodel below.)
Otherwise,
let $CFmuls$ be the set of all $\fmul{x,y}$ terms whose value in $\abst{\mu}$
is different from $\abst{\mu}[x] * \abst{\mu}[y]$.
The function \SMTrefine generates a set of axioms $\Gamma'$ such that
there exists at least one element $\fmul{x,y}$ of $CFmuls$
such that
the formula
$\abst{\varphi} \land \bigwedge \Gamma'$ has no model $\abst{\mu'}$
that agrees with $\abst{\mu}$ on the values of $x,y$ and $\fmul{x,y}$
(i.e. such that
$\abst{\mu'}[\fmul{x,y}] = \abst{\mu}[\fmul{x,y}]$,
$\abst{\mu'}[x] = \abst{\mu}[x]$ and
$\abst{\mu'}[y] = \abst{\mu}[y]$).
Intuitively, the axioms $\Gamma'$ block the bad model values for $\fmul{x,y}$,
making the abstraction more precise by restricting the set of spurious solutions.

In our current implementation, two kinds of lemmas are generated during refinement:
\emph{tangent lemmas} and \emph{monotonicity lemmas}.

\paragraph{Tangent Lemmas.}
We use the model values $\abst{\mu}[\fmul{x,y}]$, $\abst{\mu}[x]$ and
$\abst{\mu}[y]$ and \eqref{eq:mult-tangent-plane} to generate tangent
plane lemmas for $\fmul{x,y}$:
\begin{align}
\begin{split}
& \fmul{a,y} = a * y ~~\land~~ \fmul{x,b} = b * x ~~\land\\
&(((x > a \wedge y < b) \vee (x < a \wedge y > b)) \rightarrow \fmul{x,y} <  \tmul{x,y}{a,b})  ~~\land\\
&(((x <  a \wedge y < b) \vee (x >  a \wedge y >  b)) \rightarrow \fmul{x,y} >  \tmul{x,y}{a,b})\\
\end{split}
\label{eq:tangent-lemma}
\end{align}
where we can choose $a$ and $b$ as:%
\begin{align}
&a \defas \abst{\mu}[x] \text{ and } b \defas \abst{\mu}[y] \label{eq:tangent-ab1}\\
&a \defas \frac{1}{\abst{\mu}[\fmul{x,y}]} \text{ and } b \defas \abst{\mu}[y] \label{eq:tangent-ab2}\\
&a \defas \abst{\mu}[x] \text{ and } b \defas \frac{1}{\abst{\mu}[\fmul{x,y}]} \label{eq:tangent-ab3}.
\end{align}
Basically the equalities in the tangent lemma are providing
multiplication lines that enforce the correct value of \fmul{x,y} when
$x=a$ or $y=b$. Moreover, the inequalities of the tangent lemma are
providing bounds for \fmul{x,y} when $x$ and $y$ are not on the
multiplication lines.

\paragraph{Monotonicity Lemmas.}
Let $\fmul{x,y}$ and $\fmul{w,z}$ be two terms in $\abst{\varphi}$, such that
$\abs{\abst{\mu}[x]} \leq \abs{\abst{\mu}[w]}$,
$\abs{\abst{\mu}[y]} \leq \abs{\abst{\mu}[z]}$, and
$\abs{\abst{\mu}[\fmul{x,y}]} > \abs{\abst{\mu}[\fmul{w,z}]}$.
Then, we add the monotonicity lemma
\begin{equation}
  (abs(x) \leq abs(w) \wedge abs(y) \leq abs(z)) \rightarrow abs(\fmul{x,y}) \leq abs(\fmul{w,z}),
\label{eq:refinement-axioms-end}
\end{equation}
where $abs(t)$ stands for $\SMTite(t < 0, -t, t)$.

\subsubsection{Finding Models.}
\label{sec:smt-model-finding-heuristics}

It is easy to see that our algorithm
 is expected to perform much better
for unsatisfiable instances than for satisfiable ones.
The algorithm can return true (meaning that the formula is satisfiable)
only if the \UFLRA solver ``guesses'' a model
that is consistent with all the nonlinear multiplications.
In an infinite and dense domain like the reals,
the chances that this will happen are close to zero in general.

Moreover, our approach is inherently limited,
because it can only find models over the rationals.
If the input formula is satisfiable,
but all its models contain some irrational values,
then our algorithm will always abort
(or never terminate, if there is no resource budget set).
In practice, it is very likely that the same will happen
even for formulas admitting a rational solution.

One possibility for addressing this limitation would be to couple our procedure
with a complete solver for \NRA, to be used for detecting satisfiable cases,
in order to implement a more effective version of \SMTgetNRAmodel.
One such possibility is shown in Fig.~\ref{fig:smt-get-nra-model-1},
where we extract the truth assignment \abst{\psi} induced by
the \UFLRA model \abst{\mu} on the atoms of \abst{\varphi}:
\begin{eqnarray}
  \label{eq:inducedassignment}
\abst{\psi} \defas
\bigwedge_{[\abst{a_i}\in \atomsof{\abst{\varphi}}\ s.t. \abst{\mu}\models
  \abst{a_i}]} \abst{a_i} \ \ \wedge \ \
\bigwedge_{[\abst{a_i}\in \atomsof{\abst{\varphi}}\ s.t. \abst{\mu}\not\models
  \abst{a_i}]} \neg \abst{a_i},
\end{eqnarray}
We concretize it by replacing each $\fmul{x,y}$ in $\abst{\psi}$ with $x*y$,
and invoke
the complete \NRA theory solver on the resulting conjunction of \NRA-literals
$\psi$,
to check whether it contains at least one solution.
Although in general the problem is expected to be simpler than the
original input formula because the Boolean structure of
$\varphi$ is disregarded, invoking a complete \NRA theory solver at each loop
iteration of \SMTNRAcheckCEGARext could be very expensive.  Moreover,
this would still require a complete \NRA theory solver, which might not
always be available.

\begin{figure}[t]
  \centering
    \begin{fmpage}{1.0\linewidth}
  \begin{small2}
    \setcounter{pseudocodecounter}{0}
      \begin{tabbing}
        $\langle${\bf bool}, model$\rangle$ \SMTgetNRAmodel($\abst{\varphi}$, $\abst{\mu}$):\\
        \pcs xxx \= xxx \= xxx \= xxx \= xxx \= xxx \= \kill
        \pcl $\abst{\psi} = \SMTgetassignment(\abst{\mu})$
\pcc{truth assignment induced by $\abst{\mu}$ on the atoms of $\abst{\varphi}$}\\
        \pcl $\psi = \SMTconcretize(\abst{\psi})$ \pcc{replace each $\fmul{x,y}$ in $\abst{\psi}$ with $x*y$}\\
       \pcl {\bf return} \SMTNRAcheck($\psi$) \pcc{check with a complete NRA solver}
      \end{tabbing}
  \end{small2}
    \end{fmpage}
  \caption{A complete procedure using an \NRA solver.
    \label{fig:smt-get-nra-model-1}}
\ \\
  \centering
    \begin{fmpage}{1.0\linewidth}
  \begin{small2}
    \setcounter{pseudocodecounter}{0}
      \begin{tabbing}
        $\langle${\bf bool}, model$\rangle$ \SMTgetNRAmodel($\abst{\varphi}$, $\abst{\mu}$):\\
        \pcs xxx \= xxx \= xxx \= xxx \= xxx \= xxx \= \kill
        \pcl $\abst{\psi} = \SMTgetassignment(\abst{\mu})$
\pcc{truth assignment induced by $\abst{\mu}$ on the atoms of $\abst{\varphi}$}\\
        \pcl $\abst{\psi}^* =
        \abst{\psi}\wedge\SMTmultiplicationlines(\abst{\psi})$
        \pcc{add multiplication-line axioms to \abst{\psi}}\\
        \pcl {\bf return} \SMTUFLRAcheck($\abst{\psi}^*$)
      \end{tabbing}
  \end{small2}
    \end{fmpage}
  \caption{An incomplete procedure using an SMT(LRA+EUF) solver.
    \label{fig:smt-get-nra-model-2}}
\end{figure}

As an alternative, we propose the procedure
outlined in Fig.~\ref{fig:smt-get-nra-model-2},
where we extract the truth assignment \abst{\psi} induced by
the \UFLRA model \abst{\mu} on the atoms of \abst{\varphi},
and we conjoin to it the {\em multiplication lines}:
\begin{eqnarray}
  \label{eq:multiplicationlines}
\abst{\psi}^*
&\ =\ &
\abst{\psi}\ \wedge
\bigwedge_{\fmul{x,y} \in Fmuls}
\left (
  \begin{array}{ll}
        (\ x = \abst{\mu}[x] \land \fmul{x,y} = \abst{\mu}[x] * y\ )\ \lor
\\
         (\ y = \abst{\mu}[y] \land \fmul{x,y} = \abst{\mu}[y] * x\ )
  \end{array}
\right ),
\end{eqnarray}
$Fmuls$ being the usual set of all $\fmul{x,y}$ terms occurring in $\abst{\varphi}$.

The main idea is to build an LRA+EUF underapproximation $\abst{\psi}^*$ of the
NRA formula $\psi$ of Fig.~\ref{fig:smt-get-nra-model-1}, in which all
multiplications are forced to be linear.~%
Compared to the previous solution, this has the advantage of
requiring a complete SMT(LRA+EUF) solver rather than a (much more expensive)
complete \NRA solver.
  Moreover, given the simplicity of
the Boolean structure of the underapproximated formula, the check
should in general be very cheap. %
The drawback is that this is
(clearly) still an incomplete procedure.  However, in our experiments
(for which we refer to Sec.~\ref{sec:experiments}) we have found it to be
surprisingly effective for many problems.

Unlike with the basic implementation of
\SMTgetNRAmodel{}
which considers
  only one single candidate
  model at a time, the implementations in
  Fig.\ref{fig:smt-get-nra-model-1} and
  Fig.~\ref{fig:smt-get-nra-model-2} consider an
  infinite amount of them, drastically increasing the chances of
  finding a model.

\paragraph{Correctness and progress.}
  We notice that the procedure in Fig.~\ref{fig:smt-nra-pseudocode} is
  {\em correct}. In fact, it returns false only if $\varphi$ is
  NRA-unsatisfiable because by construction $\abst{\varphi}$ is  an
  over-approximation of $\varphi$, and all axioms in $\Gamma$
  are valid in any theory interpreting \fmul{x,y} as $x * y$.
  Also, it returns true only if $\vi$ is NRA-satisfiable:

\begin{itemize}
\item
  if \SMTgetNRAmodel{}
  is based only on evaluation,
  then by construction
  $\mu$ is an LRA+EUF-model for
  $\abst{\varphi}$ s.t. each $\fmul{x,y}$ equals $x*y$ in $\mu$,
  so that $\mu$ is also a model for $\varphi$;

\item  if \SMTgetNRAmodel{} is as in Fig.~\ref{fig:smt-get-nra-model-1},
then $\mu$ is an NRA-model of a conjunction of literals $\psi$ which
tautologically entails $\vi$, so that $\mu$ is a model for $\vi$;

\item  if \SMTgetNRAmodel{} is as in Fig.~\ref{fig:smt-get-nra-model-2},
then $\mu$ is an LRA+EUF-model of a conjunction of literals
$\abst{\psi}^*$ which
tautologically entails $\abst{\vi}$ and it is s.t. each $\fmul{x,y}$
equals $x*y$ in $\mu$, so that $\mu$ is a also model for $\vi$.
  \end{itemize}

We also notice that the progress of the procedure in
Fig.~\ref{fig:smt-nra-pseudocode} is guaranteed by the refinement step,
which rules out significant parts of the search space at every loop by
means of the added lemmas.

\subsubsection{Important Heuristics for Refinement.}
\label{sec:refinement-heuristics}

The description of \SMTrefine provided above leaves some flexibility
in deciding what axioms to add (and how many of them) at each iteration.
It is possible to conceive strategies with an increasing degree of eagerness,
from very lazy (e.g. adding only a single axiom per iteration)
to more aggressive ones.
In our current implementation, we eagerly add all the axioms \eqref{eq:tangent-lemma}--\eqref{eq:refinement-axioms-end}
that are violated by the current abstract solution $\abst{\mu}$,
leaving the investigation of alternative strategies as future work.
However, we found the following two strategies to be crucial for performance.

\paragraph{Tangent lemma frontiers.}
The tangent lemmas of \eqref{eq:tangent-lemma} for a given point $(a, b)$
are based on the fact that the multiplication function $x*y$ is a hyperbolic paraboloid surface,
and a tangent plane to such surface cuts the
surface into four regions such that in two of the regions the tangent plane
is above the surface, whereas in the other two regions the tangent
plane is below the surface (see Fig.~\ref{fig:plots-mult-tangent}).
Each instantiation of \eqref{eq:tangent-lemma} for a given point, therefore,
can only provide either a lower or an upper bound for a given region.
In some cases, this might lead to an infinite refinement loop
in which at each iteration the ``wrong'' bound is refined.
In order to address the problem, we use the following strategy.
For each $\fmul{x,y}$ in the input formula, we maintain a \emph{frontier} $\mktuple{l_x, u_x, l_y, u_y}$
with the invariant that whenever $x$ is in the interval $[l_x, u_x]$
or $y$ is in the interval $[l_y, u_y]$,
then $\fmul{x,y}$ has both an upper and a lower bound.
Initially, the frontiers are set to $\mktuple{0,0,0,0}$.
Whenever a lemma \eqref{eq:tangent-lemma} for $\fmul{x,y}$ is instantiated on a point $(a, b)$,
we generate further instantiations of \eqref{eq:tangent-lemma} and update the frontier as follows:
\begin{description}
\item[case $a < l_x$ and $b < l_y$:] instantiate \eqref{eq:tangent-lemma}
  on $(a, u_y)$ and on $(u_x, b)$, and set the frontier to $\mktuple{a, u_x, b, u_y}$;
\item[case $a < l_x$ and $b > u_y$:] instantiate \eqref{eq:tangent-lemma}
  on $(a, l_y)$ and on $(u_x, b)$, and set the frontier to $\mktuple{a, u_x, l_y, b}$;
\item[case $a > u_x$ and $b > u_y$:] instantiate \eqref{eq:tangent-lemma}
  on $(a, l_y)$ and on $(l_x, b)$, and set the frontier to $\mktuple{l_x, a, l_y, b}$;
\item[case $a > u_x$ and $b < l_y$:] instantiate \eqref{eq:tangent-lemma}
  on $(a, u_y)$ and on $(l_x, b)$, and set the frontier to $\mktuple{l_x, a, b, u_y}$.
\end{description}

\begin{figure}[t]
  \centering
  \begin{scriptsize}
    \begin{tabularx}{\textwidth}{cXcXcX}
      \includegraphics[scale=.1]{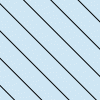} & regions with lower bounds &
     \includegraphics[scale=.1]{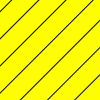} & regions with upper bounds &
     \includegraphics[scale=.1]{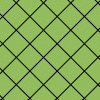} & regions with both upper and lower bounds
    \end{tabularx}
  \end{scriptsize}
~
\subfloat[current frontier]{\includegraphics[scale=.14]{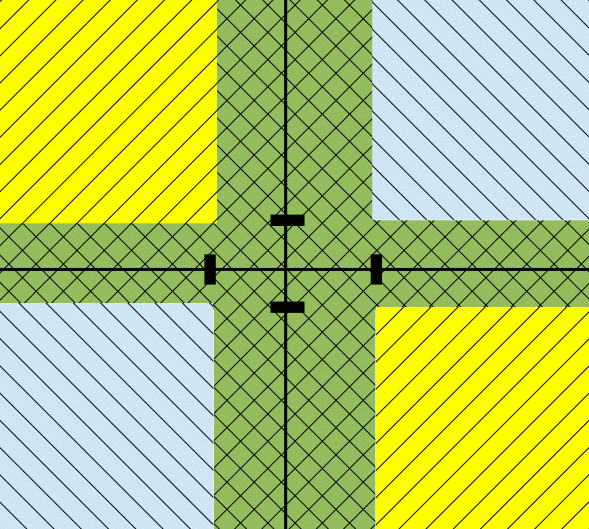}%
}~
\subfloat[new point $(a, b)$]{\includegraphics[scale=.14]{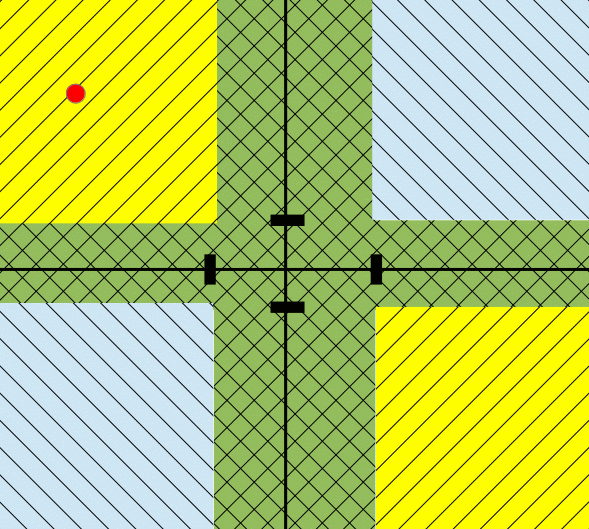}%
}~
\subfloat[instantiation of \eqref{eq:tangent-lemma} on $(a, b)$]{\includegraphics[scale=.14]{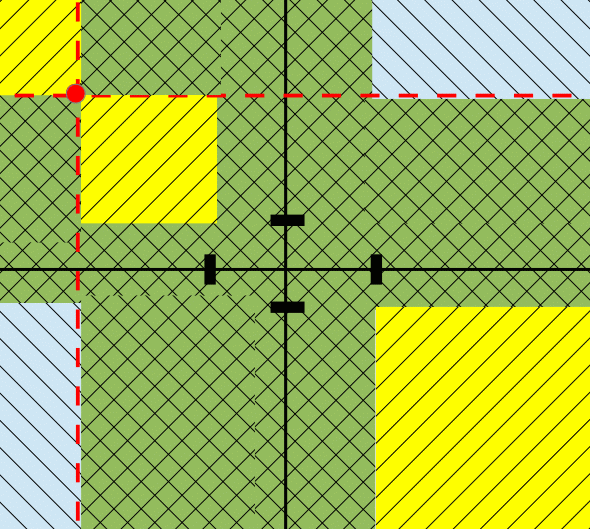}%
}~
\subfloat[additional instantiations and updated frontier]{\includegraphics[scale=.14]{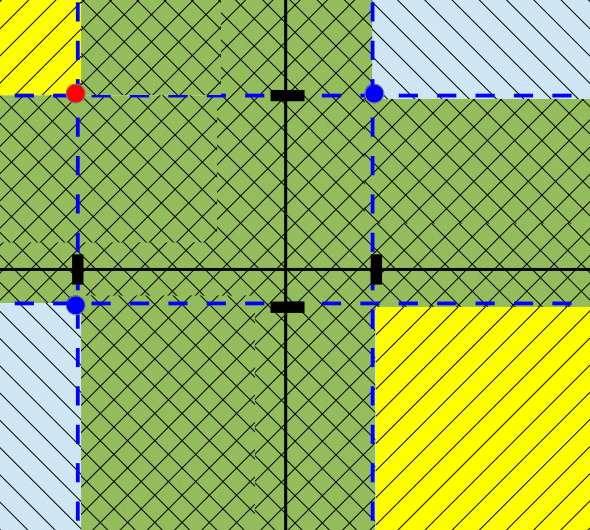}%
}
  \caption{Illustration of the tangent lemma frontier strategy.
    \label{fig:tangent-lemma-frontier}}
\end{figure}

\noindent
Fig.~\ref{fig:tangent-lemma-frontier} shows a graphical illustration of the strategy.

\paragraph{Tangent lemma rounding.}

The instantiation of a tangent lemma at the point $(a, b)$
has the side-effect of adding the rational constants $a$, $b$ and $a*b$
to the formula that is solved by the \UFLRA solver.
If such values have large numerators and/or denominators,
they might be a source of a significant slow-down for the \LRA solver
(which works on exact, arbitrary-precision rational arithmetic).
We address this issue by observing that,
in order to block a bad model $\abst{\mu}$ such that
$\abst{\mu}[\fmul{x,y}] \neq \abst{\mu}[x] * \abst{\mu}[y]$,
it is sufficient to add one of the two equalities of \eqref{eq:tangent-lemma};
therefore, instead of instantiating a tangent lemma at $(a, b)$,
we can instantiate it at either $(a+\delta, b)$ or at $(a, b+\delta)$,
for any value of $\delta$.
In practice, if $a$ (resp. $b$) is a rational constant with a very large numerator or denominator,
instead of instantiating a tangent lemma at $(a, b)$,
we instantiate two tangent lemmas at $(\lfloor a \rfloor, b)$ and $(\lceil a \rceil, b)$.

\section{From Satisfiability to Verification}
\label{sec:cegar-nra}

We now move from satisfiability checking to verification.

\subsubsection{Overview.}

In principle, the solver described in the previous section could be integrated
as a ``black box'' in any off-the-shelf SMT-based verification algorithm,
such as BMC, k-induction, or one of the many extensions of IC3 to the SMT case (e.g. \cite{ic3-ia,gpdr,spacer,ctigar}).
In practice, however, such black-box integration would hardly be effective,
especially in the case of state-of-the-art algorithms like IC3.
IC3 requires a very incremental interaction with the underlying SMT engine,
which is asked to solve a large number of relatively-cheap queries.
The procedure of Sec.~\ref{sec:cegar-nra-smt}, however, can be very expensive, especially for satisfiable queries,
which are very common in an IC3-like algorithm.%
\footnote{In fact, as already discussed in Sec.~\ref{sec:cegar-nra-smt},
the procedure is biased towards unsatisfiable instances,
and might easily diverge on satisfiable ones.}
Moreover, some of the IC3 extensions mentioned above require the ability
of performing (approximated) quantifier eliminations,
a functionality not provided by the algorithm of Fig.~\ref{fig:smt-nra-pseudocode}.

We propose therefore a white-box integration,
in which we lift the abstraction refinement approach of
Sec.~\ref{sec:cegar-nra-smt} at the transition system level.
We generate an abstract \UFLRA version of the input \NRA transition system,
which is then checked with the IC3-based procedure of \cite{ic3-ia}.
In case a counterexample is produced,
we use the \SMTNRAcheckCEGARext algorithm of Fig.~\ref{fig:smt-nra-pseudocode}
to check whether it is spurious.
If so, the axioms generated by \SMTNRAcheckCEGARext
are then used to refine the abstraction of the transition system.
The pseudo-code of this algorithm is reported in Fig.~\ref{fig:vmt-nra-pseudocode}.
Similarly to the satisfiability checking case, the $\VMTinitialabstraction$ function
replaces every non-linear multiplication $x*y$ in the input transition system and property with a $\fmul{x,y}$ term,
and adds some simple axioms about the behaviour of multiplication
 to the initial-state and transition-relation formulas of the transition system (see Sec.~\ref{sec:smt-initial-abstraction}).
In the rest of this section,
we describe the abstraction refinement algorithm in more detail.

\begin{figure}[t]
  \centering
    \begin{fmpage}{1.0\linewidth}
  \begin{small2}
    \setcounter{pseudocodecounter}{0}
      \begin{tabbing}
        {\bf bool} \VMTNRAcheck($\TS$ : transition system $\mktuple{X,I,T}$, $\varphi$ : invarant property):\\
        \pcs xxx \= xxx \= xxx \= xxx \= xxx \= xxx \= \kill
        \pcl $\abst{\TS}, \abst{\varphi} = \VMTinitialabstraction(\TS, \varphi)$\\
        \pcl {\bf while} true:\\
        \pcl \> {\bf if} budget-exhausted(): {\bf abort}\\
        \pcl \> ok, $\abst{\pi} = \VMTUFLRAcheck(\abst{\TS}, \abst{\varphi})$\\
        \pcl \> {\bf if} ok: {\bf return} true \pcc{property proved}\\
        \pcl \> $\psi = \VMTgetcexformula(\abst{\TS}, \abst{\varphi}, \abst{\pi})$ \\
        \pcl \> is\_cex, $\Gamma$ = $\SMTNRAcheckCEGARext(\psi)$\\
        \pcl \> {\bf if} is\_cex: {\bf return} false \pcc{counterexample found}\\
        \pcl \> {\bf else}: $\abst{\TS} = \VMTrefine(\abst{\TS}, \Gamma)$
      \end{tabbing}
  \end{small2}
    \end{fmpage}
  \caption{Verification of \NRA transition systems via abstraction to \UFLRA.
    \label{fig:vmt-nra-pseudocode}}
\end{figure}

\subsubsection{Counterexample Checking and Refinement.}

\begin{figure}[t]
  \centering
    \begin{fmpage}{1.0\linewidth}
  \begin{small2}
    \setcounter{pseudocodecounter}{0}
      \begin{tabbing}
        transition system \VMTrefine($\abst{\TS}$ : transition system, $\Gamma$ : set of axioms):\\
        \pcs xxx \= xxx \= xxx \= xxx \= xxx \= xxx \= \kill
        \pcl {\bf let} $\mktuple{X,\abst{I},\abst{T}} = \abst{\TS}$\\
        \pcl $\Gamma_I, \Gamma_T = \emptyset, \emptyset$\\
        \pcl {\bf for each} $\gamma$ in $\Gamma$:\\
        \pcl \> {\bf if} $\varsof{\gamma} \subseteq \attime{X}{0}$:\\
        \pcl \> \> $\Gamma_I = \Gamma_I \cup \{\mksubst{\gamma}{\attime{X}{0}}{X}\}$ \\
        \pcl \> {\bf else if} there exists $i>0$ s.t. $\varsof{\gamma} \subseteq \attime{X}{i}$: \\
        \pcl \> \> $\Gamma_T = \Gamma_T \cup \{\mksubst{\gamma}{\attime{X}{i}}{X}, \mksubst{\gamma}{\attime{X}{i}}{X'}\}$ \\
        \pcl \> {\bf else} \\
        \pcll{code:vmt-refine-multi-step-begin} \> \> {\bf let} $i$ be the smallest index s.t. $\varsof{\gamma} \cap \attime{X}{i} \neq \emptyset$ \\
        \pcll{code:vmt-refine-multi-step-end} \> \> $\Gamma_T = \Gamma_T \cup \{\mksubst{\mksubst{\gamma}{\attime{X}{i}}{X}}{\attime{X}{i+1} \cup \ldots \cup \attime{X}{i+k}}{X'}\}$ \\
        \pcll{code:vmt-refine-return} {\bf return} $\mktuple{X, \abst{I} \land \bigwedge \Gamma_I, \abst{T} \land \bigwedge \Gamma_T}$
      \end{tabbing}
  \end{small2}
    \end{fmpage}
  \caption{Refinement of the \UFLRA transition system.
    \label{fig:vmt-refine}}
\end{figure}

When \VMTUFLRAcheck returns a counterexample trace $\abst{\pi}$
for the abstract system $\abst{\TS}$,
we use \SMTNRAcheckCEGARext to check for its spuriousness.
The function \VMTgetcexformula builds a formula $\psi$ to feed to \SMTNRAcheckCEGARext,
whose unsatisfiability implies that $\abst{\pi}$ is spurious.
The formula $\psi$ is built by unrolling the transition relation of $\abst{\TS}$,
and optionally adding constraints that restrict the allowed transitions
to be compatible with the states in $\abst{\pi}$.
Various heuristics are possible, trading generality for complexity:
$\psi$ could be fully constrained by the states in $\abst{\pi}$ (thus checking only one abstract counterexample path per iteration);
it could be only partially constrained
(e.g. by considering only the Boolean variables and/or the state variables occurring only in linear constraints);
or it could be left unconstrained,
considering only the length of the abstract counterexample.
In our current implementation (see Sec.~\ref{sec:experiments}),
we use the last option,
i.e. we only consider the length of $\abst{\pi}$
to build a BMC formula that checks for any counterexample of the given length,
leaving the investigation of alternative strategies to future work.

If \SMTNRAcheckCEGARext returns true, the property is violated.
In this case, we can use the model found by \SMTNRAcheckCEGARext
to build a counterexample trace for the input system and property.

If \SMTNRAcheckCEGARext returns false,
we use the axioms $\Gamma$ produced during search to refine the transition system $\abst{\TS}$,
using the procedure shown in Fig.~\ref{fig:vmt-refine}.
Essentially, \VMTrefine translates back the axioms from their unrolled version (i.e. on variables $\attime{X}{0}, \attime{X}{1}, \ldots$) to their ``single step'' version
(on variables $X$ and $X'$), adding each of them
either to the initial-states formula or to the transition relation formula.
In case an axiom $\gamma$ spans more than a single transition step (lines \ref{code:vmt-refine-multi-step-begin}--\ref{code:vmt-refine-multi-step-end}
of Fig.~\ref{fig:vmt-refine}),
we arbitrarily choose to map the variables with the lowest index as current state variables $X$, and all the others as next-state variables $X'$.
Notice that this might cause some refinement failure, as discussed in the next paragraph.

\paragraph{Reducing the number of axioms to add.}

In general, not all the axioms generated during a call to \SMTNRAcheckCEGARext
are needed to successfully block a counterexample,
especially if eager strategies like those described in Sec.~\ref{sec:cegar-nra-smt} are used.
In the long run, having a large number of redundant axioms can be quite harmful for performance.
In order to mitigate this problem,
we apply a filtering strategy (based on unsatisfiable cores)
to the set of axioms, before adding them to the transition system.
Instead of adding $\Gamma_I$ and $\Gamma_T$ directly to $\abst{\TS}$,
we invoke the function shown in Fig.~\ref{fig:vmt-reduce-axioms}.
Note that due to the flattening of multi-step axioms described above
(lines \ref{code:vmt-refine-multi-step-begin}--\ref{code:vmt-refine-multi-step-end}
of Fig.~\ref{fig:vmt-refine}), the refinement might fail.
In this case, our current implementation simply aborts the execution.%
\footnote{We remark however that so far we have never observed this behaviour during our experiments.}

\begin{figure}[t]
  \centering
    \begin{fmpage}{1.0\linewidth}
  \begin{small2}
    \setcounter{pseudocodecounter}{0}
      \begin{tabbing}
        transition system \VMTreduceaxioms($\mktuple{X,\abst{I},\abst{T}}$, $\abst{\varphi}$, $\abst{\pi}$, $\mktuple{\Gamma_I, \Gamma_T}$):\\
        \pcs xxx \= xxx \= xxx \= xxx \= xxx \= xxx \= \kill
        \pcl $\psi = \VMTgetcexformula(\mktuple{X, \abst{I} \land \bigwedge \Gamma_I, \abst{T} \land \bigwedge \Gamma_T}, \abst{\pi})$\\
        \pcl {\bf if not} $\SMTUFLRAcheck(\psi)$:\\
        \pcl \> {\bf let} $C$ be an unsatisfiable core of $\psi$ \\
        \pcl \> $\Gamma_I = \{ \gamma \in \Gamma_I ~|~ \mksubst{\gamma}{X}{\attime{X}{0}} \in C \}$ \\
        \pcl \> $\Gamma_T = \{ \gamma \in \Gamma_T ~|~ \exists j > 0 \text{~s.t.~} \mksubst{\mksubst{\gamma}{X}{\attime{X}{j}}}{X'}{\attime{X}{j+1}} \in C \}$ \\
        \pcl \> {\bf return} $\mktuple{X, \abst{I} \land \bigwedge \Gamma_I, \abst{T} \land \bigwedge \Gamma_T}$ \\
        \pcl {\bf else} \\
        \pcl \> {\bf abort} \pcc{refinement failure}
      \end{tabbing}
  \end{small2}
    \end{fmpage}
  \caption{Reducing the axioms needed for refinement.
    \label{fig:vmt-reduce-axioms}}
\end{figure}

\section{Experimental Analysis}
\label{sec:experiments}

\newcommand{\myparagraph}[1]{\noindent\textit{#1}}

\subsubsection{Implementation and comparisons.}
We have implemented a prototype of the \VMTNRAcheck procedure 
using
the \icthreeia 
engine of \nuxmv~\cite{nuxmv} for \VMTUFLRAcheck.
The code is written in Python, 
using the \pysmt library~\cite{gario2015pysmt}.
Our implementation, benchmarks, and experimental data are available at
\url{https://es-static.fbk.eu/people/griggio/papers/tacas17-ic3-nra.tar.gz}.
We have used the following tools for our evaluation.

\myparagraph{\staticaxioms:}
we apply the upfront abstraction of \NRA to \LRA proposed in~\cite{ChampionGKT16},
running the \icthreeia engine of \nuxmv on the resulting transition system.

\myparagraph{\nrabmc-\{\zthree, \dreal\} and  \nrakind-\{\zthree, \dreal\}:}
we have implemented the \bmc~\cite{bmc} and 
\kinduction~\cite{k-induction} algorithms in Python (using \pysmt),
using either \zthree(\NRA) or \dreal(\NRA) as back-end SMT solver.

\myparagraph{\isatdefault and \isatprecision:}
we have used the latest version of the \isatthree solver~\cite{isat-hvc},
which combines an SMT solver integrating CDCL and interval constraint propagation techniques
with an interpolation-based abstaction/refinement algorithm for verification.
\isatthree supports both transition systems and software programs encoded as
control flow graphs.
Similarly to \dreal,
\isatthree may return a ``maybe unsafe'' answer and provide a
candidate solution identifying the upper and lower bounds on the
variables. 
In the experiments,
\isatdefault is the configuration suggested by the \isatthree authors
\footnote{%
\scalebox{0.8}{\texttt{-I --use-craig-interpolation --use-cegar --cegar-abstraction-inductive}}
\scalebox{0.8}{\texttt{--interpolant-rules-mcmillan --interpolant-a-biased}}
\scalebox{0.8}{\texttt{--interpolation-offset --interpolant-offset 2}}}
and \isatprecision
is the same except that the \textit{minimum splitting width} (\texttt{msw})
parameter is set to $10^{-9}$. 
We have used a smaller value for the \texttt{msw} to get more precise answers, 
i.e. ``safe'' or ``unsafe'', as suggested in the \isatthree user manual.

\subsubsection{Benchmarks.}
We have collected a total of 114 \NRA benchmarks from various sources.

\myparagraph{Handcrafted.} 
This set contains 14 hand-written instances, 13 safe and 1 unsafe.

\myparagraph{HyComp.}
The second set contains 7 benchmarks (3 safe, 4 unsafe)
which are taken
from~\cite{cimatti2012quantifier} and converted to \NRA transition
systems using \hycomp~\cite{cimatti2015hycomp}.

\myparagraph{HYST.}
This is the biggest set, consisting of 65 benchmarks.
These are generated from the Hybrid examples that come with the
\hyst~\cite{bak2015hyst} distribution, by approximating the continuous
time by sampling at a fixed time interval. This process is done
automatically using an extended version of \hyst. Since the generated
benchmarks are approximations, we do not know their safety status. The
benchmarks contain mostly non-linear behaviour.

\myparagraph{iSAT3 and iSAT3-CFG.}
The 11 benchmarks in this set (7 safe, 4 unsafe)
are taken from~\cite{isat-hvc} and the \isatthree examples available online. 

\myparagraph{nuXmv.}
In this set, we have 2 safe benchmarks which we collected from the
nuXmv users' mailing list. %
These benchmarks have complex boolean structure.

\myparagraph{SAS13.}
These 13 benchmarks are generated from the C programs used in~\cite{brain2013interpolation}, 
but interpreted over NRA instead of the theory of IEEE floating-point numbers. 
This makes some of the instances unsafe.

\myparagraph{TCM.}
We have generated 2 safe benchmarks from the Simulink models (taken
from the case study~\cite{brat2015verifying}) by first generating the
C code using the Embedded
Coder\footnote{\url{https://www.mathworks.com/products/embedded-coder/}}
and then encoding the program into a symbolic transition system.

\subsubsection{Results.} 

\begin{figure}[t]
\begin{center}
\includegraphics[width=0.5\textwidth]{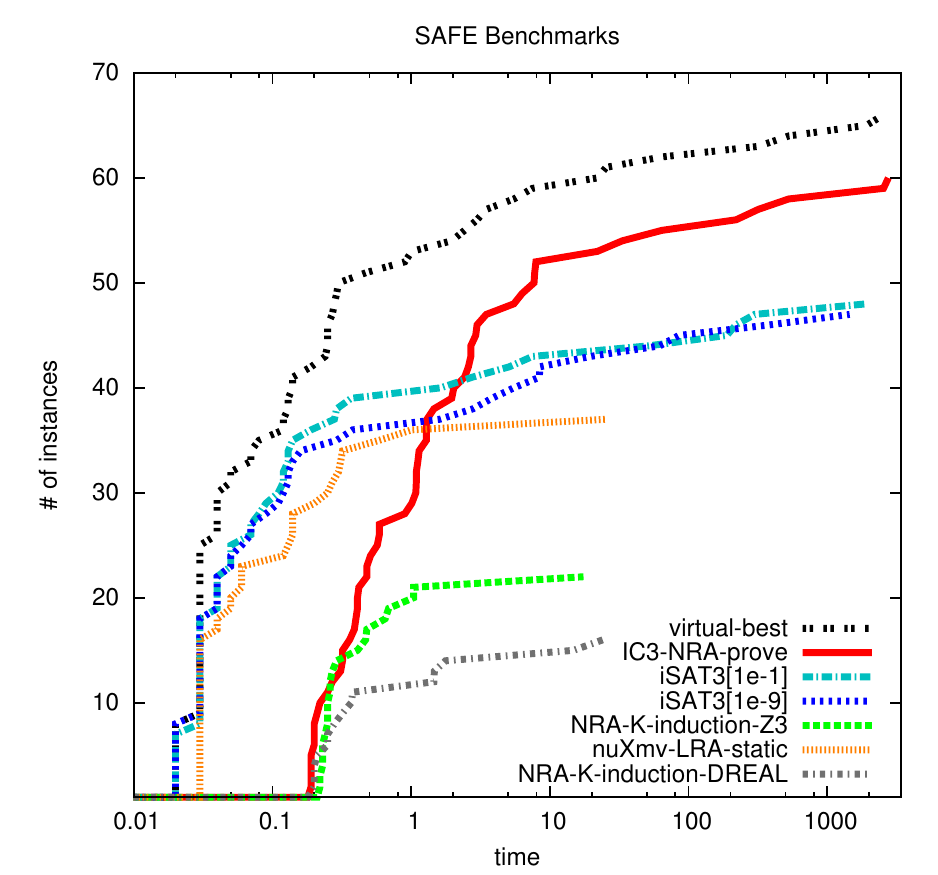}%
\includegraphics[width=0.5\textwidth]{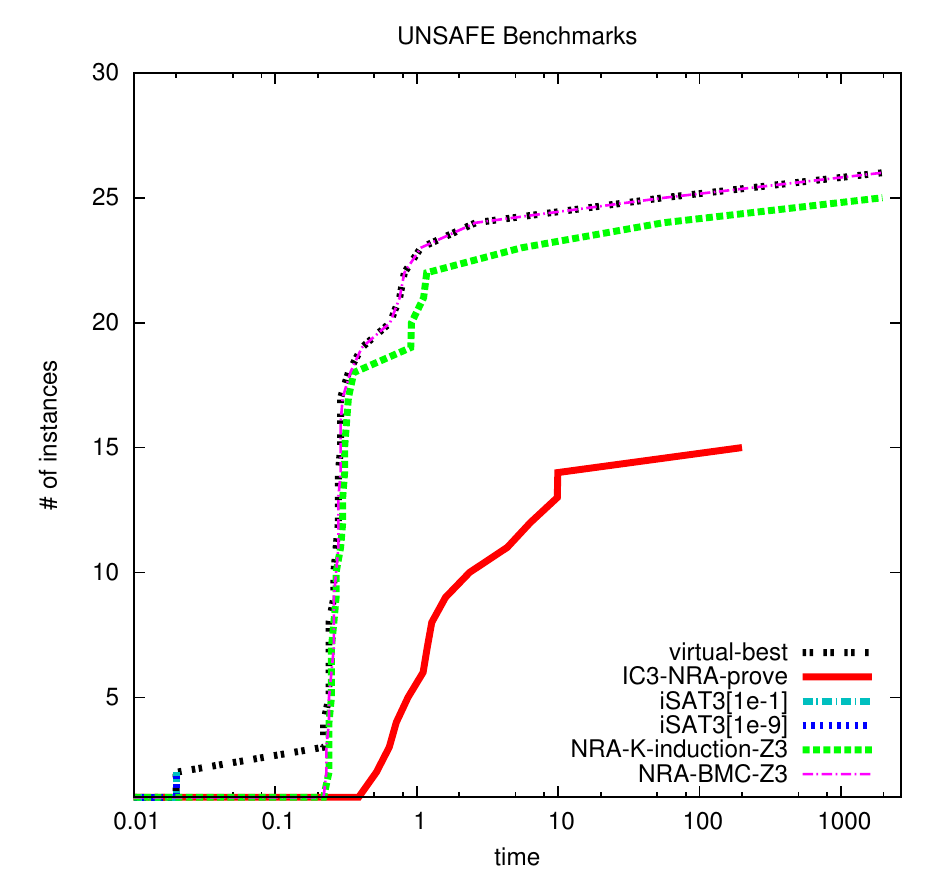}
\end{center}
\caption{Plots of SAFE and UNSAFE results.\label{fig:plots}}
\end{figure}

\begin{table}[t]
\caption{Summary of experimental results.\label{tab:results-family}}
\begin{small2}
\begin{center}
\newcommand{\ctitle}[1]{\rotatebox{-90}{\textbf{#1}~}}
\begin{tabular}{l | c | c | c | c | c | c | c | c | c }
                            & \ctitle{Total} & \ctitle{Handcrafted~}& \ctitle{HyComp~}& \ctitle{HYST~}    & \ctitle{ISAT3~}& \ctitle{ISAT3-CFG}& \ctitle{nuXmv}& \ctitle{SAS13}        & \ctitle{TCM}\\
                            & \textbf{(114)} & \textbf{(14)}        & \textbf{(7)}    & \textbf{(65)}     & \textbf{(1)}   & \textbf{(10)}     & \textbf{(2)}  & \textbf{(13)}         & \textbf{(2)}\\
\hline
\textbf{\ourapproach}       & \textbf{60}/15 & \textbf{9}/\textbf{1}& \textbf{3}/0    & 33/7              & 0/0            & \textbf{6}/2      & \textbf{2}/0  & \textbf{5}/\textbf{5} & \textbf{2}/0         \\
\textbf{\isatdefault}       & 48/2(47)       & 2/0(8)               & 0/0(3)          & \textbf{34}/2(23) & 0/0            & \textbf{6}/0(4)   & 0/0           & 4/0(9)                & \textbf{2}/0         \\
\textbf{\isatprecision}     & 47/2(19)       & 2/0(3)               & 0/0(2)          & 32/2(3)           & 0/0            & \textbf{6}/0(3)   & 0/0           & \textbf{5}/0(8)       & \textbf{2}/0         \\
\textbf{\kindzthree}        & 22/25          & 2/\textbf{1}         & 0/\textbf{2}    & 12/\textbf{15}    & 0/0            & \textbf{6}/2      & 0/0           & 0/\textbf{5}          & \textbf{2}/0         \\
\textbf{\staticaxioms}      & 37/0           & 4/0                  & 1/0             & 19/0              & 0/0            & 4/0               & \textbf{2}/0  & \textbf{5}/0          & \textbf{2}/0         \\
\textbf{\bmczthree}         & 0/\textbf{26}  & 0/\textbf{1}         & 0/\textbf{2}    & 0/\textbf{15}     & 0/0            & 0/\textbf{3}      & 0/0           & 0/\textbf{5}          & 0/0                  \\
\textbf{\kinddreal}         & 16/0(32)       & 2/0(4)               & 0/0(2)          & 9/0(19)           & 0/0            & 5/0(2)            & 0/0           & 0/0(5)                & 0/0                  \\
\textbf{\bmcdreal}          & 0/0(39)        & 0/0(8)               & 0/0(2)          & 0/0(19)           & 0/0            & 0/0(3)            & 0/0           & 0/0(7)                & 0/0                  \\
\hline
\textbf{virtual-best}       & 66/26          & 9/1                  & 3/2             & 38/15             & 0/0            & 7/3               & 2/0           & 5/5                   & 2/0                  \\
\hline
\end{tabular}
\end{center}

Each column shows a benchmark family, and each entry gives the number of safe/unsafe instances found. 
For tools working over interval arithmetic, the number of ``maybe unsafe'' is reported in parentheses.
\end{small2}
\end{table}
\begin{table}[t]
\begin{center}
\caption{Comparitive summary of total solved
  benchmarks.\label{tab:results-gain-loss}}
\begin{small2}
\newcolumntype{C}[1]{>{\centering}m{#1}}
\begin{tabular}{ l | c | c | C{8em} | c | c }
                             & \textbf{\# Solved} & \textbf{\# Uniquely Solved} & \textbf{Difference wrt. \ourapproach}  & \textbf{Gained} & \textbf{Lost} \\
\hline
\hline
\textbf{\ourapproach}        & \textbf{60}/15  & \textbf{9}/0             & -                              & -              & -             \\
\hline
\textbf{\staticaxioms}       & 37/0            & 0/0                      & -38                            & 1/0            & 24/15         \\
\hline
\textbf{\isatdefault}        & 48/2(47)        & \multirow{2}{*}{4/0}     & -25                            & 4/0            & 16/13         \\
\textbf{\isatprecision}      & 47/2(19)        &                          & -26                            & 3/0            & 16/13         \\
\hline
\textbf{\kindzthree}         & 22/25           & \multirow{2}{*}{0/\textbf{11}} & -28                            & 2/11           & 40/1          \\
\textbf{\bmczthree}          & 0/\textbf{26}   &                          & -49                            & 0/11           & 60/0          \\
\hline
\textbf{\kinddreal}          & 16/2(32)        & \multirow{2}{*}{0/0}     & -59                            & 2/0            & 46/15         \\
\textbf{\bmcdreal}           & 0/0(39)         &                          & -75                            & 0/0            & 60/15         \\
\hline
\hline
\textbf{virtual-best}        & 66/26           & -                        & 17                             & 6/11           & 0             \\
\hline
\end{tabular}
\end{small2}
\end{center}
\end{table}

We ran our experiments on a cluster of machines with 2.67GHz Xeon
X5650 CPUs and 96GB of RAM, running Scientific Linux 6.7. We used 6GB
memory limit and 3600 seconds CPU timeout.

The results are summarized in Tables~\ref{tab:results-family} and \ref{tab:results-gain-loss} 
and in Fig.~\ref{fig:plots}. 
The plots show 
the time to solve an
instance on the x-axis and the total number of solved
instances on the y-axis. 
Table~\ref{tab:results-family} reports a summary of the solved instances by family,
whereas
Table~\ref{tab:results-gain-loss} shows
a comparitive analysis by reporting for each tool the number of uniquely solved instances and the difference of solved instances w.r.t
\ourapproach.
We can make the
following observations from the experimental results:
\begin{itemize}
\item \ourapproach is the best performer overall,
  and it significantly outperforms all the other approaches on safe instances (where it can solve 9 problems that are out of reach for all the other tools).
Interestingly, despite its simplicity, 
our model finding approach (as outlined in \sref{sec:cegar-nra-smt})
is surprisingly effective, allowing \ourapproach to find 15 counterexample traces.
\item The simple abstraction proposed in \cite{ChampionGKT16} is quite effective for many families,
allowing \staticaxioms to verify more properties than the approaches based on K-induction with an \NRA solver.
However, \ourapproach results in a clear and very significant improvement, 
solving more than twice as many instances than \staticaxioms (and losing only 1).
\item None of the other tools (with the exception of \staticaxioms) 
  is able to solve any safe benchmark in the HyComp and nuXmv families.
  These benchmarks have a non-trivial Boolean structure and a significant linear component. 
  Both \ourapproach and \staticaxioms are able to fully exploit the effectiveness of the underlying \icthreeia engine of \nuxmv,
  outperforming the competitors.
  However, \ourapproach is very competitive also on the HYST family, 
  whose instances are mostly non-linear and have very little Boolean structure. 
\item Increasing the default precision of \isatthree significantly reduces the number of ``maybe unsafe'' answers,
but it doesn't seem to help in solving more benchmarks.
In fact, we remark that even with the increased precision \isatprecision 
classifies 2 safe instances as ``maybe unsafe'' (whereas in the default configuration, 6 safe instances are classified as ``maybe unsafe'').
\end{itemize}

\section{Conclusions and Future Work}
\label{sec:conclusion}

We presented a novel abstraction-refinement approach to the
verification of transition systems with nonlinear dynamics expressed
in the NRA theory. We abstract non-linear multiplication as an
uninterpreted function, leveraging efficient invariant checkers for
transition systems over LRA and EUF to solve the problem in the
abstract space. In case of spurious counterexample, the abstraction of
multiplication is incrementally refined by introducing suitable
axioms, based on the idea of tangent planes.
An extensive experimental evaluation demonstrates that the proposed
approach is significantly more effective than approaches directly
based on SMT(NRA) solving.

This work opens up several important directions.
First, we are going to improve the implementation, by integrating all
the steps within the \nuxmv~\cite{nuxmv} model checker, and to perform
a thorough analysis of the various heuristic choices.
Second, we will investigate the potential of the approach for
SMT, both for other theories (e.g. NIA) and for extended functionalities
(e.g. interpolation).
We will also extend the scope of the approach to deal with
transcendental functions, look-up tables, and partially axiomatized
functions (e.g. gain functions known to be monotonic and of restricted
co-domain).

Finally, we are going to investigate the generalization of the
approach from transition systems to continuous-time hybrid systems
with nonlinear characteristic functions.

\paragraph{Acknowledgement.}
We greatly thank the \isatthree team for providing the latest
\isatthree executable and iSAT3-CFG benchmarks. We also thank James
Davenport for the fruitful discussions on CAD techniques
and finding solutions in NRA.

\newpage
\bibliographystyle{plain}
\bibliography{ref}

\end{document}